\documentclass[a4paper, 11pt]{article}\usepackage[]{graphicx}\usepackage[]{color}

\usepackage{alltt}
\usepackage[left=3cm,right=3cm, top = 2.8cm, bottom = 2.8cm]{geometry}

\newif\ifbiom
\biomfalse 

\usepackage{natbib}
\usepackage{url}
\usepackage[utf8]{inputenc}
\usepackage[UKenglish]{babel}
\usepackage[T1]{fontenc}
\usepackage{amsmath}
\usepackage{amssymb}
\usepackage{color}

\newcommand{\given}{\,\vert\,}

\newcommand{\eg}{\textit{e.g.}\ }
\newcommand{\ie}{\textit{i.e.}\ }
\newcommand{\simind}{\mathrel{\overset{\text{ind}}{\thicksim}}}

\DeclareMathOperator{\Var}{Var} 

\usepackage{xr}
\IfFileExists{upquote.sty}{\usepackage{upquote}}{}
\begin{document}

\title{A marginal moment matching approach for fitting endemic-epidemic models to underreported disease surveillance counts}

\ifbiom
\else
\author{Johannes Bracher and Leonhard Held}
\fi

\ifbiom
\else
\maketitle 
\fi

\begin{abstract}
Count data are often subject to underreporting, especially in infectious disease surveillance. We propose an approximate maximum likelihood method to fit count time series models from the endemic-epidemic class to underreported data. The approach is based on marginal moment matching where underreported processes are approximated through completely observed processes from the same class. Moreover, the form of the bias when underreporting is ignored or taken into account via multiplication factors is analysed. Notably, we show that this leads to a downward bias in model-based estimates of the effective reproductive number. A marginal moment matching approach can also be used to account for reporting intervals which are longer than the mean serial interval of a disease. The good performance of the proposed methodology is demonstrated in simulation studies. An extension to time-varying parameters and reporting probabilities is discussed and applied in a case study on weekly rotavirus gastroenteritis counts in Berlin, Germany.
\end{abstract}

\section{Introduction}
\label{sec:introduction}

Data from routine surveillance of infectious diseases rarely reflect the entire burden in a population \citep{Gibbons2014, Noufaily2019}. At the community level, not all infected individuals seek medical care, \eg due to absence or mildness of symptoms, which is referred to as under-ascertainment. At the healthcare level, part of those seeking healthcare are not diagnosed correctly, lack lab confirmation or are not entered in the notification system. This is often referred to as underreporting, but we will use the term in a broader sense so that it also subsumes under-ascertainment. Various study types exist to estimate the degree of underreporting, including community-based survey studies \citep{Woudenberg2019}, returning traveller studies, capture-recapture methods and serological surveys \citep{Gibbons2014}. For common and permanently immunizing childhood diseases, birth numbers can be used to infer reporting probabilities \citep{Becker2017}.

Underreporting has received a fair amount of attention in infectious disease modelling and statistics. Approaches of varying complexity have been suggested, starting from simple multiplication factor methods (\eg \citealt{Jandarov2014}, \citealt{Stocks2018}). More sophisticated techniques have been described for branching processes \citep{Fraser2009, White2010,Azmon2014} and compartmental models \citep{Cauchemez2008, Dorigatti2012, Gamado2014, Fintzi2017, Magal2018}
. Moreover, disease mapping approaches for underreported spatial \citep{Bailey2005, Oliveira2017} and spatio-temporal \citep{Stoner2019, Sharmin2018a} count data have been suggested. Underreported count outcomes in regression have been considered by \cite{Dvorzak2016}. In statistical count time series modelling, however, underreporting is rarely addressed. Exceptions are \citeauthor{Fernandez-Fontelo2016} (\citeyear{Fernandez-Fontelo2016}, \citeyear{Fernandez-Fontelo2019}) who suggested a hidden integer valued autoregressive (INAR) model and \cite{Sharmin2018} who use a hidden log-linear autoregressive count model.
Here we examine the impact of underreporting in the endemic-epidemic (EE) modelling framework \citep{Held2005,Held2012,Meyer2014,Meyer2017} which has links to integer-valued generalized autoregressive conditional heteroscedasticity (INGARCH) models (\citealt{Ferland2006}, \citealt{Zhu2011}). This model class is of particular interest in epidemiological applications as it can be derived from a mechanistic model of disease transmission \citep{Wakefield2019} and serve to estimate effective reproductive numbers. The EE class has been used for this purpose by \cite{Wang2011} and \cite{Bauer2018}, but without accounting for underreporting of surveillance data.

Our main contribution consists in an approximate maximum likelihood method to fit EE models to underreported data. It is based on a marginal moment matching approach where underreported processes are approximated by completely observed processes from the same class. The reporting probability needs to be specified based on external knowledge, as it is generally not identifiable from time series data alone. The sensitivity of the estimates to the chosen value can subsequently be examined. The arguments underlying the likelihood approximation also have implications on the biases occurring when underreporting is ignored or taken into account using multiplication factors. Avoiding such biases is especially relevant when EE models are not just used for prediction, but to estimate effective reproductive numbers.

A marginal moment matching approach can also be used to fit models which are defined at a finer time scale than the observed data. Even if data are only available for weekly reporting intervals one can then fit models defined at a half-weekly time scale, which is often desirable from an epidemiological perspective. An overview of similar temporal aggregation problems in classical time series analysis is given by \cite{Silvestrini2008} while \cite{Braennaes2002} address an application to count time series.

The article is structured as follows. In Section \ref{sec:endemic_epidemic} we describe an EE model, extend it to account for underreporting and derive the marginal moments of the underreported process. We also consider the impact of temporal aggregation of observations
on the marginal moments. These steps form the basis of the approximate inference scheme introduced in Section \ref{sec:inference} and assessed in simulation studies in Section \ref{sec:simulations}. In Section \ref{sec:real_data} we extend the approach to models with time-varying parameters and apply it in a case study on rotavirus gastroenteritis in Berlin, Germany. Section \ref{sec:discussion} concludes with a discussion.

\section{Methodology}
\label{sec:endemic_epidemic}

\subsection{Model definition}
\label{subsec:model_definition}

The EE modelling framework for time series of infectious disease counts was introduced by \cite{Held2005} and extended in a series of articles (\citealt{Paul2008,Held2012,Meyer2014,Meyer2017}). While much of the existing work treats multivariate models, we here focus on the univariate version. We start by assuming all model parameters to be time-constant but consider time-varying parameters in Section \ref{subsec:model_specification_application} and Web Appendix A.

Conditional on the past, the number $X_t$ of cases of a given disease in time period $t = 1, 2, \dots$ is then assumed to follow a negative binomial distribution
\begin{align}
X_t \given \mathbf{X}_{<t} & \sim \text{NegBin}(\lambda_t, \psi)\label{eq:X_t}
\end{align}
with conditional mean $\lambda_t$ and overdispersion parameter $\psi > 0$ so that $\Var(X_{t} \given \mathbf{X}_{<t}) = \lambda_{t} + \psi\lambda_{t}^2$. Here we use $\mathbf{X}_{<t}$ to denote the history $X_{t-1}, \dots, X_1$. The negative binomial distribution allows the model to adapt to the typically high dispersion in surveillance data. In the simplest case the conditional expectation $\lambda_t$ is additively decomposed into
\begin{align}
\lambda_t & = \nu + \phi X_{t - 1} \label{eq:cond_mean_simple}.
\end{align}
Here the term $\nu > 0$ is interpreted as the \textit{endemic component} of incidence, \ie the part which cannot be explained by previous cases. This can \eg represent infection from sources outside the study region or indirect transmission from environmental sources.
The autoregressive term $\phi X_{t - 1}$ with $\phi \geq 0$, also called the \textit{epidemic component}, captures the dynamics of the process, see the next section for a more detailed discussion.

In \cite{Bracher2019} we recently relaxed the AR(1) assumption from \eqref{eq:cond_mean_simple} by regressing on a weighted sum of past observations. The weights represent the serial interval distribution, \ie the distribution of time between onset of clinical symptoms in a case and a secondary case infected by the first \citep{Becker2015}. A related approach is to include an autoregression on $\lambda_{t - 1}$, as is done in INGARCH models \citep{Ferland2006, Fokianos2016}:
\begin{align}
X_t \given \mathbf{X}_{<t}, \lambda_1 & \sim \text{NegBin}(\lambda_t, \psi)\label{eq:X_t2}\\
\lambda_t & = \nu + \phi X_{t - 1} + \kappa\lambda_{t - 1}, \label{eq:mean_feedback}
\end{align}
where $\kappa \geq 0$. The initial value $\lambda_1$ is assumed to be random (a common assumption is that it follows the respective stationary distribution). This corresponds to a negative binomial INGARCH(1, 1) model \citep{Zhu2011}. We will work with this extended definition, which will turn out to be particularly useful in the context of underreporting.

In the following we consider the case where the process $\{X_t\}$ is not fully observable. Instead, we observe an underreported process $\{\tilde{X}_t\}$ which depends on $\{X_t\}$ through independent binomial thinning,
\begin{equation}
\tilde{X}_{t} \given {X}_{t} \simind \text{Bin}(X_{t}, \pi),\label{eq:reporting_step}
\end{equation}
where $0 < \pi \leq 1$ is the reporting probability.

\subsection{Effective reproductive number and serial interval distribution}
\label{subsec:Reff}

\cite{Bauer2018} recently derived the AR(1) version \eqref{eq:X_t}--\eqref{eq:cond_mean_simple} of the EE model from a time-discrete SIR (susceptible-infected-removed) model with a large population and a relatively rare disease. They showed that if the serial interval corresponds to one time step, the parameter $\phi$ can be interpreted as the local effective reproductive number $R_{\text{eff}}$, \ie the average number of secondary cases an infected individual causes in the considered population, given all partial immunities and intervention measures. This approach had previously been used by \cite{Wang2011}.

For the extended model \eqref{eq:X_t2}--\eqref{eq:mean_feedback} the conditional mean $\lambda_t$ can be re-written as (\citealt{Fokianos2016}, p.7)
\begin{equation}
\lambda_t = \sum_{d = 0}^{t - 2} \kappa^{d} \nu \ + \ \frac{\phi}{1 - \kappa} \cdot \sum_{d = 1}^{t - 1} (1 - \kappa)\kappa^{d - 1} X_{t - d} \ + \ \kappa^{t - 1}\lambda_{1}, \label{eq:mean_seasonal_geometric}
\end{equation}
which has the form of an \textit{epidemic renewal equation} \citep[S17]{Fraser2009}. The effective reproductive number is thus
\begin{equation*}
R_{\text{eff}} = \frac{\phi}{1 - \kappa},
\end{equation*}
while the serial interval distribution is assumed to be geometric with parameter $(1 - \kappa)$. This corresponds to a mean serial interval of $1/(1 - \kappa)$ at the time scale of the observation interval. While there is often no specific biological mechanism that would justify a geometric serial interval distribution, it represents as a decrease of infectiousness over time.

\subsection{Marginal moments of the underreported process}
\label{subsec:properties}

As shown by \citeauthor{Zhu2011} (\citeyear{Zhu2011}), $\{X_t\}$ is second-order stationary if $(\phi + \kappa)^2 + \phi^2\psi < 1$, with marginal mean, variance and autocorrelation function
\begin{align}
\mu = \frac{\nu}{1 - \xi}, \ \ \sigma^2 = \frac{1 - \xi^2 + \phi^2}{1 - \xi^2 - \psi\phi^2} \cdot (\mu + \psi\mu^2) \label{eq:moments_unthinned}, \ \ \mbox{ and }  \rho(d)  = \eta \xi^{d - 1},\ \ d = 1, 2, \dots ,
\end{align}
respectively. Here we denote by
\begin{equation}
\eta = \phi \cdot \frac{1 - \kappa\xi}{1 - \xi^2 + \phi^2} \label{eq:g}\ \ \ \text{and} \ \ \  \xi = \phi + \kappa
\end{equation}
the first-order autocorrelation and the decay factor of the autocorrelation function, respectively.

The corresponding quantities for the underreported process $\{\tilde{X}_t\}$ are
\begin{align}
\tilde{\mu} = \pi\mu, \ \ \tilde{\sigma}^2 = \pi^2\sigma^2 + \pi(1 - \pi)\mu,\label{eq:moments_thinned} \ \ \mbox{ and }  \tilde{\rho}(d) = \tilde{\tau} \eta\xi^{d - 1}, \  d = 1, 2, \dots,
\end{align}
where the {\em attenuation factor}
$\tilde{\tau} = \tilde{\rho}(d)/\rho(d)$
can be written as
\begin{equation*}
\tilde{\tau} = 1 - (1 - \pi) \cdot \frac{\tilde{\mu}}{\tilde{\sigma}^2},
\end{equation*}
see Web Appendix B.1 for the derivation of \eqref{eq:moments_thinned}. The attenuation of the autocorrelations gets stronger, \ie $\tilde{\tau}$ gets smaller, as $\pi$ and the index of dispersion $\tilde{\sigma}^2/\tilde{\mu}$ decrease.

\subsection{Second-order equivalent models}
\label{subsec:second_order_equivalence}

An interesting implication of equations \eqref{eq:moments_unthinned}--\eqref{eq:moments_thinned} is that different combinations of reporting probabilities and parameters for the latent process can lead to the same marginal second-order properties of the underreported process.
In fact, for any underreported process $\{\tilde{X}_t\}$ with underlying parameters $\nu, \phi, \kappa, \psi$ and reporting probability $\pi$, and any $\pi_Y\in (\pi, 1]$, one can find a process
\begin{align}
Y_t \given \mathbf{Y}_{<t}, \lambda_{Y, 1} & \sim \text{NegBin}(\lambda_{Y, t}, \psi_Y) \nonumber\\
\lambda_{Y, t} & = \nu_{Y} + \phi_Y Y_{t - 1} + \kappa_Y \lambda_{Y, t -1 } \label{eq:lambda_Y}
\end{align}
so that the underreported process
$$
\tilde{Y}_t \given Y_t \simind \text{Bin}(Y_t, \pi_Y)
$$
has the same marginal second-order properties as $\{\tilde{X}_t\}$:
\begin{align}\label{eq:M3}
\tilde{\mu}_Y & = \tilde{\mu}; \ \ \ \tilde{\sigma}^2_Y = \tilde{\sigma}^2; \ \ \ \tilde{\rho}_Y(d) = \tilde{\rho}(d) \text{ for } d = 1, 2, \dots
\end{align}
We then call $\{\tilde{X}_t\}$ and $\{\tilde{Y}_t\}$ \textit{second-order equivalent}. While they are not equal in distribution, they behave very similarly, see the simulation study in Web Appendix E.1. An important special case emerges for $\pi_Y = 1$ when $\{\tilde{X}_t\}$ and the fully observed process $\{{Y}_t\}$ are second-order equivalent.

The equality of marginal means, variances and autocorrelation functions in \eqref{eq:M3} implies a system of four equations, which yields explicit formulas for the parameters of the process $\{{Y}_t\}$:
\begin{align}
\nu_Y & = \frac{\tilde{\mu}(1 - \xi)}{\pi_Y} = \frac{\pi}{\pi_Y} \cdot \nu \label{eq:nu_Y},\\
\phi_Y & = \frac{\sqrt{\tilde{\tau}^2_Y(1 - \xi^2)^2 + 4(\tilde{\tau}_Y \xi - \tilde{\tau}\eta)\tilde{\tau}\eta(1 - \xi^2)} - \tilde{\tau}_Y(1 - \xi^2)}{2(\tilde{\tau}_Y\xi - \tilde{\tau}\eta)} \label{eq:phi_Y},\\
\kappa_Y & = \xi - \phi_Y\label{eq:kappa_Y},\\
\text{and} \ \ \ \psi_Y & = \frac{(1 - \xi^2)\tilde{\tau}_Y \tilde{\sigma}^2 - \pi_Y \tilde{\mu}\{1 - \xi^2 + \phi_Y^2\}}{\phi_Y^2\tilde{\tau}_Y \tilde{\sigma}^2 + \tilde{\mu}^2 \{1 - \xi^2 + \phi_Y^2\}} \label{eq:psi_Y},
\end{align}
where
$\tilde{\tau}_Y = 1 - (1 - \pi_Y)\tilde{\mu}_Y/\tilde{\sigma}_Y^2 = 1 - (1 - \pi_Y)\tilde{\mu}/\tilde{\sigma}^2$.
Derivations are given in Web Appendix B.2, where it is also shown that for $\pi_Y \in (\pi, 1]$ the parameters $\nu_Y$, $\phi_Y$, $\kappa_Y$, and $\psi_Y$ are all non-negative, \ie $\{Y_t\}$ is a well-defined process. The equations \eqref{eq:nu_Y}--\eqref{eq:psi_Y} simplify in the
important special case $\pi_Y = 1$, see equations (W20)--(W23) in Web Appendix B.2. It can also be shown that $\nu_Y$ and $\phi_Y$ are monotonically decreasing in $\pi_Y$, while $\kappa_Y$ and $\psi_Y$ are monotonically increasing (Web Appendix B.3). This has implications on the biases occurring when underreporting is ignored, see Section \ref{subsec:bias_naive_analyses}. The presented argument on second-order equivalence of processes with different reporting probabilities can even be extended to models with time-varying parameters (Section \ref{subsec:model_specification_application} and Web Appendix A).

\subsection{Marginal moments of temporally aggregated processes}
\label{subsec:coarsened}

An assumption underlying the arguments in Section \ref{subsec:Reff} is that the serial interval distribution is well described by a geometric distribution at the time scale of the observation intervals, typically weeks. For diseases with serial intervals of a few days, however, it would be more reasonable to define the model $\{X_t\}$ at a half-weekly time scale, say. Ignoring underreporting for the moment, we can then only observe the \textit{temporally aggregated} process $\{X^*_t\}$ where
\begin{equation}
X^*_t = X_{2t - 1} + X_{2t}. \label{eq:coarsening_step}
\end{equation}
The mean, variance and autocorrelation function of $\{X^*_t\}$ can be shown to be
\begin{equation}
\mu^* = 2\mu; \ \ \ \ \ \sigma^{*2} = 2(1 + \eta)\sigma^2; \ \ \ \ \ \rho^*(d) =  \eta \cdot \frac{(1 + \xi)^2}{2(1 + \eta)}\cdot \xi^{2(d - 1)},
\label{eq:moments_coarsened}
\end{equation}
where $\eta$ and $\xi$ are defined in \eqref{eq:g}.
The autocorrelation function of $\{X^*_t\}$ is thus of the same general form as that of $\{X_t\}$ given in \eqref{eq:moments_unthinned}. Similar results for various classical time series models can be found in \cite{Silvestrini2008}. Furthermore, if $\kappa \geq 2 - \sqrt{3} \approx 0.27$, which for half-weekly time steps corresponds to a serial interval of at least $4.8$ days, one can find a process $\{Y_t\}$ from our class which is defined directly at the time scale of $\{X^*_t\}$ and shares its marginal second-order properties. Details and expressions for the parameters of $\{Y_t\}$ can be found in Web Appendix C.2.

The same second-order equivalence argument can be made for processes $\{\tilde{X}^*_t\}$ which are both underreported and temporally aggregated. Interestingly, a well-defined second-order equivalent process $\{Y_t\}$ which is fully observed and defined at the time scale of $\{\tilde{X}^*_t\}$ can then even be found in many cases where $\kappa < 0.27$. This is a consequence of the correlation attenuation caused by underreporting. As notation gets somewhat involved, details on processes with both underreporting and temporal aggregation have been moved to Web Appendix C.3.

\section{Inference}
\label{sec:inference}

\subsection{Need for external information on the reporting probability}

An implication of the results from Section \ref{subsec:second_order_equivalence} is that it is generally not feasible to estimate the reporting probability from time series data alone. The reason is that practically identical fits can be achieved with quasi-equivalent models featuring different reporting probabilities, see also the simulation study in Section \ref{subsec:comparison_forward_algorithm}. In order to estimate $\pi$ one would need to fix one of the other parameters. Similar identifiability issues have been reported for count data regression models \citep{Dvorzak2016} and latent INAR time series models \citep{Bracher2019a}. For compartmental models, simulation studies have led to related conclusions. For example, \citeauthor{Fraser2009}(\citeyear{Fraser2009}, S18) state that the reporting probability and epidemic parameters cannot be estimated jointly.

\subsection{Na\"ive approaches and their biases}
\label{subsec:bias_naive_analyses}

In practical analyses of surveillance data underreporting is often ignored, but within the EE class this will lead to biased parameter estimates. Assume that the data come from a process $\{\tilde{X}_t\}$ with reporting probability $\pi$ and underlying parameters $\nu, \phi, \kappa, \psi$. If we wrongly assume full reporting, the parameter estimates can be expected to be around the parameters of a process $\{Y_t\}$ which is second-order equivalent to $\{\tilde{X}_t\}$, but has reporting probability $\pi_Y = 1$, see equations (W20)--(W23) in Web Appendix B.2. The derivatives $\text{d}\nu_Y/\text{d}\pi_Y$ etc. given in Web Appendix B.3 imply that this leads to overestimation of $\kappa$ and $\psi$ and underestimation of $\nu$, $\phi$ and $R_{\text{eff}}$. Note that an upward bias in $\kappa$ means we overestimate the average serial interval $1/(1 - \kappa)$.

A simple approach to account for underreporting is to inflate counts with a multiplication factor, given by the inverse of the assumed reporting probability, prior to analysis (\eg, \citealt{Jandarov2014}, \citealt{Stocks2018}). This changes their overall level and variance, but leaves the autocorrelations unaffected. As $\phi$ and $\kappa$ reflect the autocorrelation structure, we do not expect this approach to alleviate bias in these parameters. We can use \eqref{eq:nu_Y}--\eqref{eq:psi_Y} to find the parameters $\nu_Y, \phi_Y, \kappa_Y, \psi_Y$ of a fully observed process $\{Y_t\}$ which is second-order equivalent to the process $\{X_t/\pi\}$. This tells us which biases to expect, see Section \ref{subsec:bias_precision} for a simulation study exploring this in more detail.

\subsection{Maximum likelihood estimation via marginal moment matching}
\label{subsec:approximate_mle}

The challenge in fitting the hierarchical model \eqref{eq:X_t2}--\eqref{eq:reporting_step} to data is that the likelihood cannot be evaluated directly. There is no obvious way of recursively computing the likelihood contributions $\text{Pr}(\tilde{X}_t = \tilde{x}_t \given \tilde{X}_{t - 1} = \tilde{x}_{t - 1}, \dots \tilde{X}_1 = \tilde{x}_1)$ as for fully observed models. Methods for hidden Markov models cannot be applied either as the latent process $\{\tilde{X}_t\}$ lacks the Markov property, see equation \eqref{eq:mean_seasonal_geometric}. Equations \eqref{eq:nu_Y}--\eqref{eq:psi_Y}, however, offer two potential solutions. Both are based on marginal moment matching where underreported processes are approximated by second-order equivalent processes without underreporting, \ie a reporting probability of one. This is inspired by a method for fitting Gaussian AR($p$) models to mismeasured data by expressing them as ARMA($p$, $p$) models \citep{Staudenmayer2005}.

A simple strategy is to fit a model with $\pi = 1$, \ie ignoring underreporting, and subsequently de-bias the parameter estimates. The corrected estimates are given by the parameters $\nu_Y, \phi_Y, \kappa_Y, \psi_Y$ underlying a process $\{\tilde{Y}_t\}$ which is second-order equivalent to the fitted model $\{X_t\}$, but underreported with the assumed reporting probability $\pi_Y$. These can be obtained using equations \eqref{eq:nu_Y}--\eqref{eq:psi_Y}. However, as $\pi_Y > \pi = 1$ does not hold here, this does not ensure that the corrected estimates for $\kappa$ and $\psi$ are non-negative. Especially if the true $\kappa$ is small this poses a problem, see the simulation study in Web Appendix E.7.

We therefore adopt a more elaborate approach where underreporting is accounted for during rather than after the likelihood optimization. In order to approximate the likelihood of a model $\{\tilde{X}_t\}$ with underlying parameters $\nu, \phi, \kappa, \psi$ and reporting probability $\pi$, we consider the second-order equivalent model $\{Y_t\}$ without underreporting ($\pi_Y = 1$) and parameters $\nu_Y, \phi_Y, \kappa_Y, \psi_Y$ as in \eqref{eq:nu_Y}--\eqref{eq:psi_Y}. Such a process always exists and is well-defined as $\pi_Y = 1 \geq \pi$. The likelihood of $\{\tilde{X}_t\}$ can then be approximated by that of $\{Y_t\}$, which is easier to evaluate. We also estimate the initial value $\lambda_{1}$ \citep{Ferland2006} and for simplicity set $\lambda_{Y, 1} = \pi\lambda_1$. We can then compute
\begin{align}
& \mathcal{L}(\nu, \phi, \kappa, \psi, \lambda_{1}, \pi) \approx \mathcal{L}(\nu_{Y}, \phi_Y, \kappa_Y, \psi_Y, \lambda_{Y, 1}, \pi_Y = 1) = \label{eq:likelihood}\\
& \Pr(Y_1 = \tilde{x}_1 \given \lambda_{Y, 1}, \psi_Y) \cdot \prod_{t = 2}^T \Pr(Y_t = \tilde{x}_t \given \nu_{Y}, \phi_Y, \kappa_Y, \psi_Y, \lambda_{Y, 1}, Y_{t - 1} = \tilde{x}_{t - 1}, \dots, Y_1 = \tilde{x}_1),\nonumber
\end{align}
where $\tilde{x}_1, \dots \tilde{x}_T$ are the observed counts. The likelihood contribution of time $t$ is a negative binomial probability mass function with mean $\lambda_{Y, t}$ and overdispersion parameter $\psi_Y$, see equation \eqref{eq:lambda_Y}. Maximization of the likelihood function for a given $\pi$ can be done using standard numerical optimization and $\pi$ can be varied to assess the sensitivity of the estimates to this choice. This procedure can be extended to models with time-varying parameters, see Web Appendix A.

\subsection{Marginal moment matching to account for temporal aggregation}
\label{sec:moment_matching_coarsening}

Using the results from Section \ref{subsec:coarsened}, a marginal moment matching approach can also be used to fit EE models to temporally aggregated data, \eg to fit a model defined at a half-weekly time scale to weekly counts. The likelihood of a model $\{X_t\}$ defined in half-weekly steps can be approximated by that of a weekly model $\{Y_t\}$ which is second-order equivalent to the temporally aggregated process $\{X^*_t = X_{2t - 1} + X_{2t}\}$, see equation \eqref{eq:moments_coarsened}. If $\kappa > 0.27$, the process $\{Y_t\}$ is always well-defined and the likelihood \eqref{eq:likelihood} can be evaluated.

If $\kappa < 2 - \sqrt{3} < 0.27$, the process $\{Y_t\}$ may not be well-defined as $\kappa_Y$ can get negative. However, it can be shown that $\kappa_Y$ cannot get smaller than $3 - \sqrt{8} \approx - 0.17$ (Web Appendix C.2). In many practical settings, such a modest negative value of $\kappa_Y$ does not hamper evaluation of the likelihood \eqref{eq:likelihood}. If one of the negative binomial likelihood contributions cannot be evaluated due to a negative conditional expectation $\lambda_{Y, t}$, we pragmatically set $\lambda_{Y, t} = \nu_{Y, t}$ in order to evaluate the likelihood.

The approximation steps to account for underreporting and temporal aggregation are straightforward to combine so that we can also fit models accounting both for underreporting and temporal aggregation. As mentioned before, the condition $\kappa > 0.27$ can even be relaxed if we assume $\pi < 1$. Details have been moved to Web Appendix C.3.

\section{Simulation studies}
\label{sec:simulations}

In the simulation studies presented here we focus on underreported, but not temporally aggregated processes. Simulation studies on accounting simultaneously for both can be found in Web Appendix E.5 and Web Appendix E.6.

\subsection{Comparison to forward algorithm}
\label{subsec:comparison_forward_algorithm}

If $\kappa = 0$ the latent process $\{X_t\}$ is a first-order Markov chain and the exact likelihood of an underreported EE model can be evaluated with the forward algorithm \citep{Zucchini2009}. This is now used to assess the quality of proposed approximation method.

We consider the agreement over a broad range of parameter combinations, independently sampling $N = 1000$ sets of parameters from uniform distributions $\nu \sim \text{Unif}(3, 30)$, $\phi \sim \text{Unif}(0.01, 0.99)$, $\psi \sim \text{Unif}(0.001, 0.2)$, $\pi \sim \text{Unif}(0.01, 1)$, $\lambda_1 \sim \text{Unif}(0.5\mu, 2\mu)$, where $\mu = \nu/(1 - \phi)$. As in the estimation method we treat the initial value $\lambda_1$ as an additional unknown parameter. Sets of parameters with $\phi^2(1 + \psi) \geq 1$, \ie which imply non-stationarity, were discarded. For each set of parameters we generated a time series of length 100 from the respective underreported process $\{\tilde{X}_t\}$ and evaluated the log-likelihood at the true parameter values using the forward algorithm and the proposed approximation method.

\begin{figure}[h!]
\includegraphics[width=\textwidth]{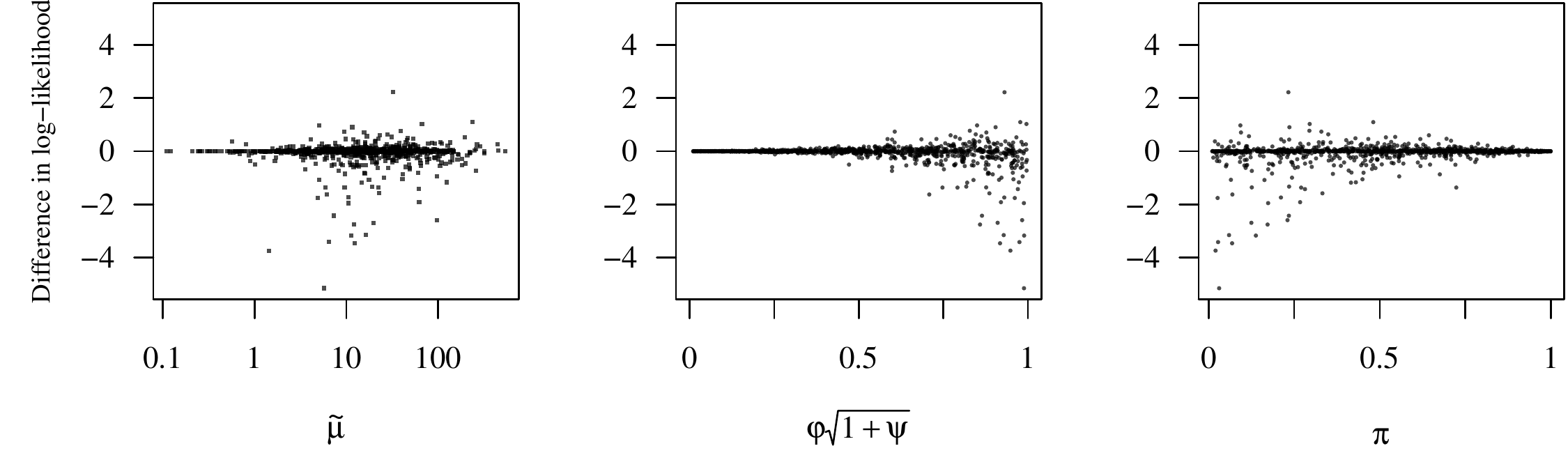} 
\caption{Difference between the log-likelihood values returned by the proposed approximation and the forward algorithm, evaluated at the true parameter values. The differences for 1000 randomly generated parameter combinations are plotted against the stationary mean $\tilde{\mu}$ of the underreported process $\{\tilde{X}_t\}$ (on a log scale, left); $\phi\sqrt{1 + \psi}$, which describes how close the process is to non-stationarity (threshold at 1, middle); the reporting probability $\pi$ (right).}
\label{fig:comparison_fp}
\end{figure}

Figure \ref{fig:comparison_fp} shows the differences between the two methods, plotted against characteristics of the generating model. The agreement between both methods is very good, with absolute differences below 0.1 in 73\% and below 1 in 97\% of the cases. The approximation works somewhat less well as $\phi\sqrt{1 + \psi}$ approaches 1, \ie close to non-stationarity (middle panel), and when the reporting probability $\pi$ is low (right panel). The computation time required for the forward algorithm was two orders of magnitude longer than for the approximation method.

The presented results also support our assertion that the reporting probability cannot be estimated from time series data. After all, the approximating processes with $\pi_Y = 1$ led to equally good fits to the data as the true generating models with reporting probability $\pi$.

We have also conducted a simulation study comparing the maximum likelihood estimates obtained via the forward algorithm and the moment matching approach in Web Appendix E.8. Again, agreement between the two is very good.

\subsection{Bias and loss of precision due to underreporting}
\label{subsec:bias_precision}

\begin{figure}[h!]
\center
\includegraphics[width=0.9\textwidth]{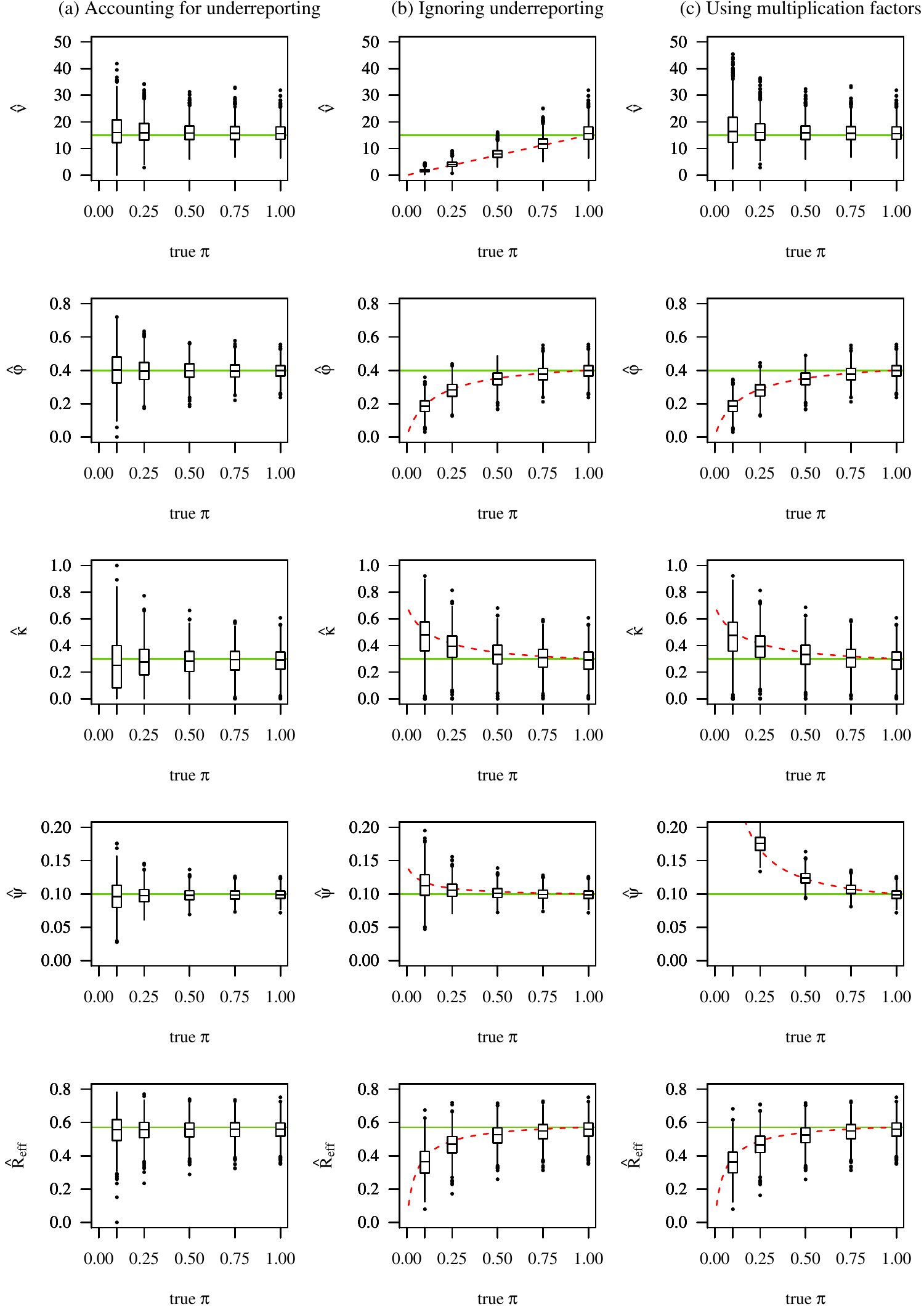} 
\caption{Empirical distribution of parameter estimates under different degrees of underreporting. (a) With the proposed method and the correct $\pi$. (b) Ignoring underreporting, \ie assuming $\pi = 1$. (c) Applying the multiplication factor $1/\pi$ prior to analysis. Solid lines show true parameter values, dashed lines expected biases.}
\label{fig:bias_variance}
\end{figure}

In a second simulation study we examine whether the arguments from Section \ref{subsec:bias_naive_analyses} describe the biases in na\"ive analyses well and whether the proposed estimation method can avoid them. We simulated 1000 sample paths of length 416 (\ie eight years of weekly data, as in the case study in Section \ref{sec:real_data}) from a model with $\nu = 15, \phi = 0.4, \kappa = 0.3$ and $\psi = 0.1$. This implies a marginal mean of $\mu = 50$ and a standard deviation of $\sigma \approx 20$, while the effective reproductive number is $R_{\text{eff}} = 0.57$. We generated underreported versions of these time series with five different reporting probabilities $\pi \in \{0.1, 0.25, 0.5, 0.75, 1\}$, where $\pi = 1$ corresponds to full reporting. Selected sample paths can be found in Web Appendix E.3. The underreported time series are analysed (a) using the proposed estimation method and the correct $\pi$, (b) ignoring underreporting and (c) inflating the counts with a factor of $1/\pi$ (with subsequent rounding) prior to fitting the model. The results are shown in Figure \ref{fig:bias_variance}.

The proposed method avoids bias with increasing variability of the estimates as $\pi$ -- and thus the data quality -- decreases. Ignoring underreporting leads to pronounced biases for low values of $\pi$. These are well described by equations \eqref{eq:nu_Y}--\eqref{eq:psi_Y} with $\pi_Y = 1$, compare Section \ref{subsec:bias_naive_analyses} (dashed lines in the plots). The multiplication factor method avoids bias in $\nu$, but not the other parameters. We re-ran the analysis with time series of length 208 (four years of weekly data) and 832 (sixteen years), which gave similar results (Web Appendix E.3).

We also assessed the coverage of Wald confidence intervals for the different parameters in Web Appendix E.4. Empirical coverage was close to the nominal level given sufficient training data. For shorter time series there was some under-coverage especially when the reporting probability was small.

We moreover performed a simulation study featuring both underreporting and temporal aggregation, using the methodology described in Sections \ref{sec:moment_matching_coarsening} and Web Appendix C.3 for fitting. The results, shown in Web Appendix E.5 indicate that the proposed methodology works well also in this case. This was confirmed in a more realistic setting where we used the fitted model from Section \ref{sec:real_data} as the data generating process, see Web Appendix E.6.

\section{Case study: rotavirus gastroenteritis in Berlin}
\label{sec:real_data}

Rotavirus causes gastroenteritis with symptoms including diarrhoea, vomiting and fever and affects mainly young children \citep{Heymann2015}. It is transmitted predominantly via the faecal oral route. \citet{Grimwood1983} report a mean serial interval of 4.9 days for children and 6.4 days for adults. Previous infection has an immunizing effect, but immunity is imperfect and waning over time. Since 2006 a vaccine has been available in Germany and since 2013 it has been recommended by the Standing Committee on Vaccination.

\begin{figure}[h!]
\center
\includegraphics[width=\textwidth]{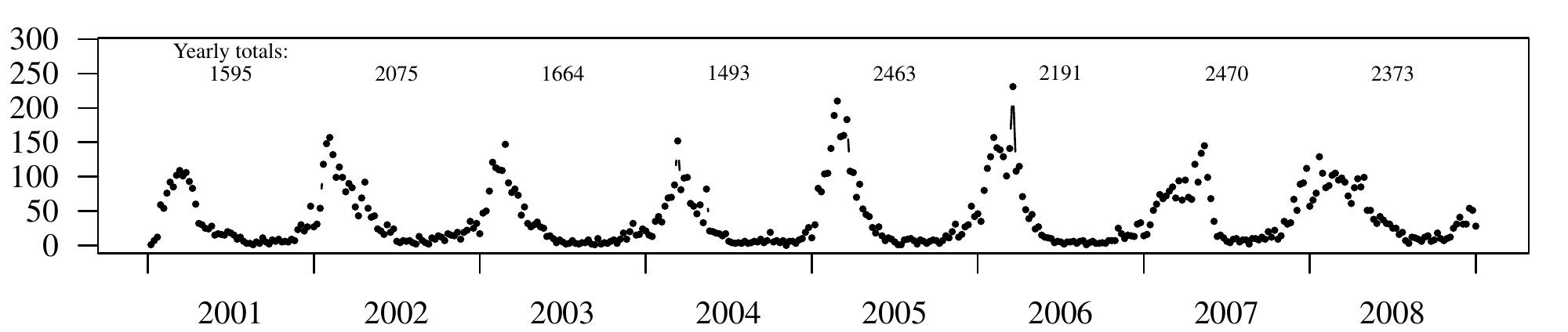} 
\caption{Weekly numbers of laboratory-confirmed rotavirus cases in Berlin, 2001--2008.}
\label{fig:data}
\end{figure}

Rotavirus is a notifiable disease in Germany. We analyse weekly counts of laboratory-confirmed cases in Berlin, provided by the Robert Koch Institute via its web platform (\texttt{\url{https://survstat.rki.de/}}). The data cover the years 2001--2008 and are displayed in Figure \ref{fig:data}. Incidence in this period has previously been assumed to be largely unaffected by vaccination, while from 2009 onwards effects of increased coverage could be observed (\citealt{Weidemann2014}). To avoid this extra level of complexity we restrict our analysis to the pre-vaccination period as was done in \cite{Stocks2018}.

Rotavirus surveillance counts are known to be strongly underreported \citep{Tam2012}. For the period 2001--2008 the reporting probability has been estimated as 4.3\% (95\% credibility interval: 3.9\% to 4.7\%) in the Western Federal States of Germany, including Berlin \citep{Weidemann2014}.

\subsection{Model specification}
\label{subsec:model_specification_application}

We apply an EE model accounting for both underreporting and temporal aggregation to the rotavirus data.
The latter is considered because the average serial interval of rotavirus is reported to be less than one week \citep{Grimwood1983}.
Rotavirus is often modelled using more complex mechanistic approaches describing the effects of births, partial immunity acquired by infection and the waning thereof (\citealt{Atkins2012, Weidemann2014, Stocks2018}). Such an approach is beneficial to inform intervention strategies, but
a considerable number of parameter values must be pre-specified to enable identifiability of the remaining ones \citep{Weidemann2014}.
Here we take a different approach, where the above factors are subsumed in the effective (rather than basic) reproductive number and its variation over the course of a season. A similar approach has been taken by \cite{Gaalen2017}. The simplicity of the EE model formulation then enables us to explicitly assess the impact of underreporting.

For the rotavirus data we extend model \eqref{eq:X_t2}--\eqref{eq:reporting_step} by allowing $\nu_t$ and $\phi_t$ to vary over time and adjust for temporal aggregation of half-weekly data, \ie
\begin{align}
X_t \given \mathbf{X}_{<t}, \lambda_1 & \sim \text{NegBin}(\lambda_t, \psi) \nonumber \\
\lambda_t & = \nu_t + \phi_t X_{t - 1} + \kappa\lambda_{t - 1}, \label{eq:model_rota}\\
\tilde{X}_t \given X_t & \simind \text{Bin}(X_t, \pi) \nonumber\\
\tilde{X}^*_t & = \tilde{X}_{2t - 1} + \tilde{X}_{2t}. \label{eq:coarsening_rota}
\end{align}
As the latent $\{X_t\}$ is defined for half-weekly time steps we set \citep{Held2012}
\begin{align*}
\log(\nu_t) & = \alpha^{(\nu)} + \gamma^{(\nu)} \sin(2\pi t/104) + \delta^{(\nu)} \cos(2\pi t/104),\\
\log(\phi_t) & = \alpha^{(\phi)} + \gamma^{(\phi)} \sin(2\pi t/104) + \delta^{(\phi)} \cos(2\pi t/104)
\end{align*}
to account for seasonal variation with a period length of 104 half-weeks. The parameter $\kappa$ (and thus the serial interval distribution) is assumed to be time-constant. Exploratory analyses indicated that including more than one pair of sine/cosine waves or linear time trends into $\log(\nu_t)$ or $\log(\phi_t)$ can lead to slight improvements in AIC, but consistently worse BIC. As the respective parameter estimates were not significantly different from zero (based on Wald confidence intervals) we did not include these additional terms.

The models are fitted using the approach described in Web Appendix A,
which extends that from Section \ref{subsec:approximate_mle} to time-varying parameters and temporally aggregated observations. Given the initial value $\lambda_1$, marginal means, variances and autocovariance functions of $\{\tilde{X}^*_t\}$ can be recursively computed even if $\nu_t$ and $\phi_t$ vary over time. The parameters of a second-order equivalent process $\{Y_t\}$ without underreporting, \ie $\pi_Y = 1$, and defined directly at the weekly time scale, can be obtained using a second recursive procedure. As in equation \eqref{eq:likelihood} the likelihood of the process $\{\tilde{X}^*_t\}$ can then be approximated by that of $\{Y_t\}$, and likelihood optimization can be done using numerical optimization.

\subsection{Results}
\label{subsec:results_rota}

\begin{figure}[h!]
\includegraphics[width=\textwidth]{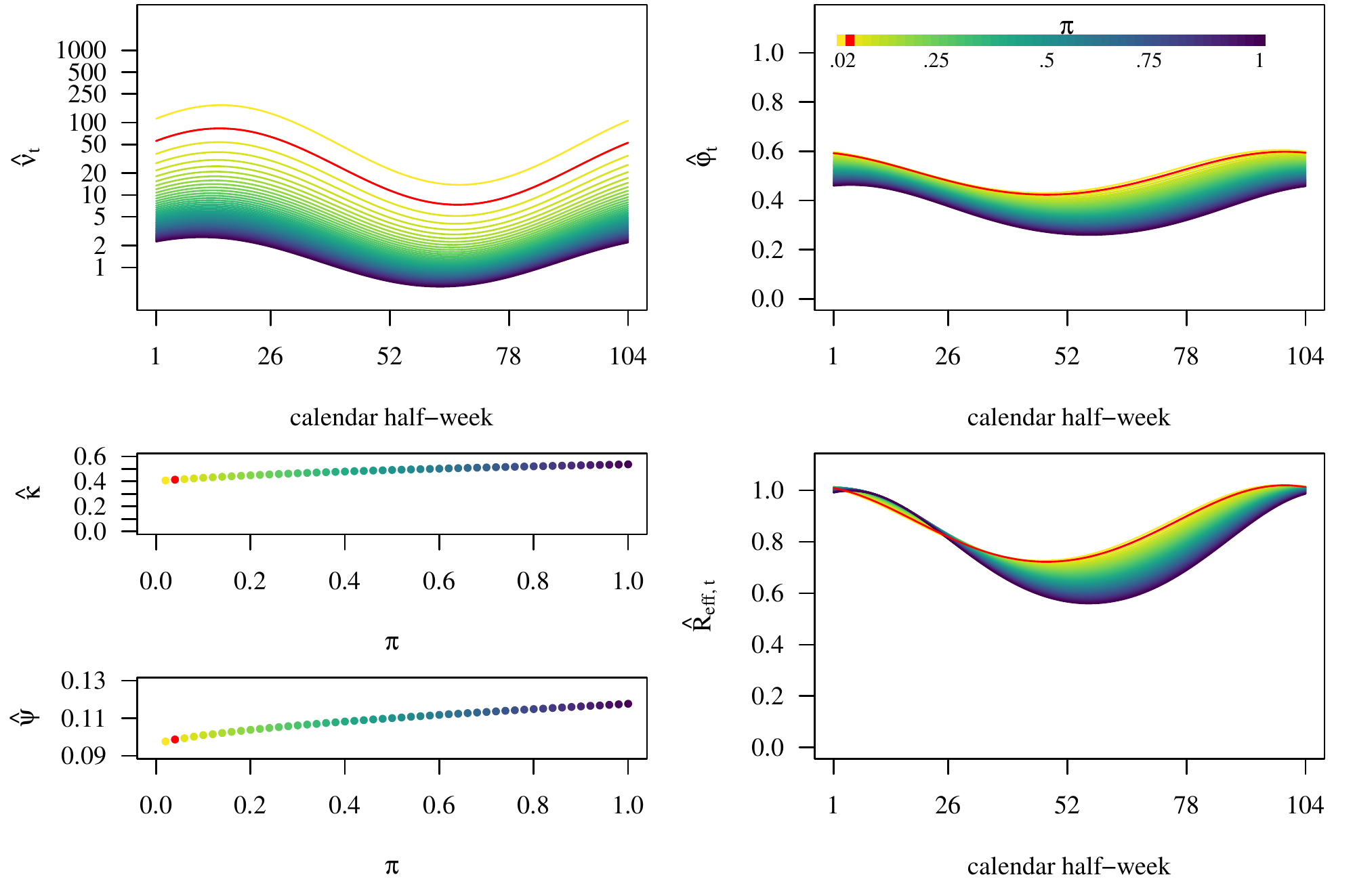} 
\caption{Parameter estimates of the underreported EE model applied to weekly rotavirus counts in Berlin, as functions of the assumed reporting probability $\pi$. The bottom right panel shows the time-varying effective reproductive number $R_{\text{eff}, t} = \phi_t/(1 - \kappa)$. The fit with $\pi = 0.043$ as in Weidemann et al. (2014) is highlighted in red. This figure appears in color in the electronic version of this article.}
\label{fig:results}
\end{figure}

We fitted model \eqref{eq:model_rota}--\eqref{eq:coarsening_rota} for different reporting probabilities $\pi \in \{0.02, 0.04, \dots, 0.98, 1\}$. Plots showing fitted and observed values as well as the residual autocorrelation function are provided in Web Appendix F.1. The parameter estimates obtained under different assumptions on the reporting probability are displayed in Figure \ref{fig:results}. The estimated effective reproductive number $R_{\text{eff}, t} = \phi_t/(1 - \kappa)$ shows considerable seasonal variation, with higher values during winter and early spring. With $\pi = 0.043$ as in \cite{Weidemann2014}, $R_{\text{eff}, t}$ ranges from 0.72 in calendar week 24 to 1.02 in calendar week 50 (highlighted in red). On average we have $R_{\text{eff}, t} < 1$, so that the endemic component $\nu_t$, representing infections imported from other regions, is necessary to maintain rotavirus transmission. Comparing the estimates of the endemic terms $\nu_t$ and marginal expectations $\mu_t$, we find that the endemic component accounts for roughly one in 10 cases. The estimated $\kappa = 0.41$ implies a mean serial interval of $1/(1 - 0.41)\times 3.5 \text{ days } = 6.0$ days, which is slightly larger than the estimate by \cite{Grimwood1983}.

Under the unrealistic assumption of no underreporting ($\pi = 1$) we obtain an estimate of $\kappa = 0.54$ and thus a longer average serial interval of 7.5 days. The seasonal range for $R_{\text{eff}}$ is now from 0.56 to 1 in this case (darkest purple line in Figure \ref{fig:results}), lower than for $\pi = 0.043$. The sensitivity of the estimates to the assumed reporting probability is thus stronger during the off-season.

\cite{Weidemann2014} state that the reporting probability may have increased in 2005 due to changed reimbursement schemes, and this is in line with the somewhat increased reported incidence in the seasons 2005--2008. We therefore also fitted a model where $\pi_t = 0.043$ in the years 2001 through 2004, then increases linearly to 0.063 over the course of the year 2005 and stays there until 2008. However, as Figure 5 and Web Table W5 show, the resulting parameter estimates are similar to those obtained with a time-constant $\pi = 0.043$.

\begin{figure}
\includegraphics[width=\textwidth]{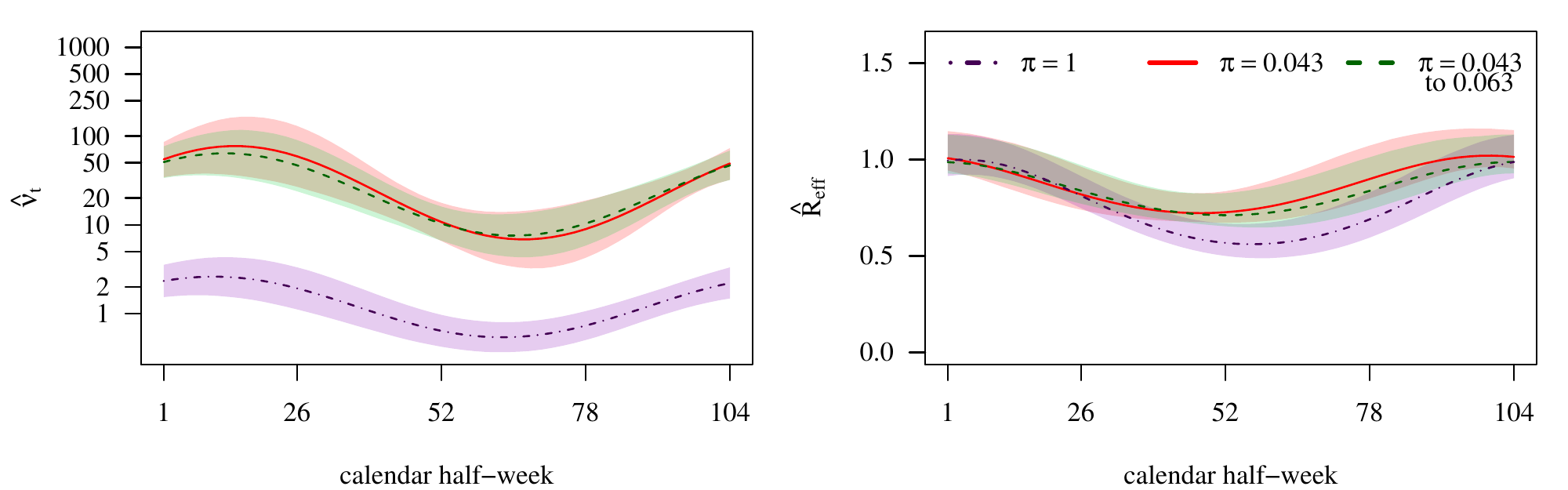} 
\caption{Estimated time-varying endemic parameters $\nu_t$ and effective reproductive numbers $R_{\text{eff}, t}$ when assuming (a) $\pi = 1$ (purple, dash-dotted line), (b) $\pi = 0.043$ (red, solid line) and (c) time-varying reporting probability increasing from $\pi_t = 0.043$ to $\pi_t = 0.063$ over the course of 2005 (green, dashed line). Point-wise 90\% confidence intervals are shown as shaded areas. This figure appears in color in the electronic version of this article.}
\label{fig:Reff_uncertainty}
\end{figure}

To gain an understanding of the uncertainty in the estimates, Figure \ref{fig:Reff_uncertainty} also shows 90\% Wald confidence intervals. As $R_{\text{eff}, t}$ is a complicated non-linear function of the estimated parameters, rather than using the delta method we computed the corresponding confidence intervals by repeatedly sampling the parameters from a multivariate normal distribution with the maximum likelihood estimates as the mean and the inverse Fisher information as the covariance matrix. This is somewhat heuristic as even for completely observed models with time-constant parameters, asymptotic normality of the conditional maximum likelihood estimators is only established if $\psi$ is known \citep{Cui2017}. However, simulation studies show that these intervals have good coverage (Web Appendix E.6).

Lastly we fitted the same model without the aggregation step \eqref{eq:coarsening_rota}, \ie defined at a weekly time scale. This led to somewhat lower values of $R_{\text{eff}}$ between 0.65 and 1.00. The results are shown in Supplementary Figure W17. The average serial interval then was 7.9 days (corresponding to $\hat{\kappa} = 0.11$), while the endemic component accounts for roughly one in six cases.

Our estimated effective reproductive numbers are lower than the estimated basic (rather than effective) reproductive numbers from \cite{Stocks2018} who obtain a seasonal range of 0.9 to 1.2 for $R_0$. In \cite{Gaalen2017}, a study based on Dutch data, the effective reproductive number shows a similar seasonal pattern as in Figure \ref{fig:results}, but is higher (between 0.85 and 1.15, Figure 3 in \citealt{Gaalen2017}). A possible explanation is that \citealt{Gaalen2017} do not allow for any infections from outside of their study region. Note also that both \cite{Stocks2018} and \cite{Gaalen2017} account for underreporting via multiplication factors.


\section{Discussion}
\label{sec:discussion}

We extended the EE modelling framework to account for underreporting. In simulation studies we showed that the proposed likelihood approximation is highly accurate and that, combined with the correct reporting probability, our inference method avoids bias. The case study on rotavirus gastroenteritis showed that estimates of effective reproductive numbers from EE models increase when underreporting is taken into account.
Accounting for temporal aggregation increases the estimated reproductive numbers further.

We modelled underreporting through binomial thinning, which is a commonly used, but simple mechanism \citep{Azmon2014,Fintzi2017,Stoner2019}. It assumes that reporting of different cases is independent, which may not always be the case. For instance, certain general practitioners or hospitals systematically fail to report their cases. More sophisticated reporting processes could in principle be added to the model, but its properties would then get more complex.

A number of alternatives to the proposed approximate maximum
likelihood method exist. Preliminary analyses with a full Bayesian
approach have been hampered by two difficulties. Firstly, mixing of standard MCMC samplers was poor. To circumvent this, more sophisticated algorithms seem to
be needed. Secondly, prior choice is challenging in EE
models \citep{Bauer2018}. Advantages, however, are that the
uncertainty about the reporting probability $\pi$ could be included
more easily and that the hidden process $\{X_t\}$ could be
reconstructed. Alternatively it may be worthwhile to combine the proposed likelihood
approximation with a hybrid Bayesian approach, allowing for uncertainty
in the reporting probability $\pi$ but preventing feedback using the cut algorithm
\citep{Plummer2015}. An alternative maximum likelihood estimation approach could be based
on particle filtering, as implemented in the
\texttt{R} package \texttt{pomp} \citep{King2016}. While this
represents a powerful general tool, it is computationally
demanding and requires careful choice of starting values to obtain
reliable results \citep{Stocks2018}.

We focused on univariate EE models, whereas much of the
previous work on the class covered spatio-temporal settings (most
recently \citealt{Meyer2017}, \citealt{Bracher2019}). It is not
entirely straightforward to extend our approach to this
case and Bayesian methods as applied in a similar context \eg by \cite{Gomez-Rubio2019}
may then be required for inference.
However, some of our qualitative findings will still be relevant. They
indicate that underreported processes are well approximated by
processes with an autoregression on $\lambda_{t - 1}$ or, as in
equation \eqref{eq:mean_seasonal_geometric}, higher-order lags. If the
goal is prediction \citep{Held2017} rather than
parameter estimation, there should be little benefit in
explicitly accounting for underreporting. Also, we observed that the
biases were weaker
when counts were high. They may thus be negligible if counts are in
the hundreds or even thousands (\eg \citealt{Bauer2018}). An
additional difficulty in multivariate settings is to account for
differences in reporting probabilities between strata, where reliable
information is rarely available.

In our application to rotavirus we used marginal moment matching to fit
a model defined for half-weekly time steps to weekly data. This approach is appropriate
for an average serial interval of roughly five days as reported by
\cite{Grimwood1983}, and indeed we obtained a quite similar estimate of six days.
In this context note that other studies have assumed shorter serial
intervals for rotavirus (\eg 2.1 days in \citealt{Gaalen2017}). Dynamics within
a half-week then play a role, and more sophisticated data augmentation
methods would likely be required to fit models under this assumption to weekly data.

Our fitted model implies that on average the effective reproductive number within Berlin is below one and infections from external sources, which account for around one in 10 cases, are necessary to maintain transmission. Given that slightly more than half a million persons commute to or from Berlin \citep{BundesagenturArbeit2020}, a city with 3.6 million inhabitants, this does not seem excessive. However, as rotavirus is predominantly spread between children, this aspect may be less relevant than for other diseases. It is possible that other attenuation phenomena are at work which cause an underestimation of effective reproductive numbers. For example, reporting delays, where cases may be registered one or several weeks after their actual occurrence (\citealt{Azmon2014}, \citealt{Noufaily2019}), may contribute to additional attenuation.

\section*{Acknowledgements}

We would like thank the Associate Editor and two anonymous reviewers for helpful comments.

\bibliographystyle{apalike}
\bibliography{Bib_underreporting_short.bib}

\section*{Data availability statement}

This article uses data publicly available from the Robert Koch Institut (\texttt{\url{https://survstat.rki.de/}}). The data are also available in the \texttt{R} package referenced in \textit{Supporting Information}.

\section*{Supporting Information}

Web Appendices, Tables, and Figures referenced in Sections this article are available at the Biometrics website on Wiley Online Library. The suggested methods have been implemented in \texttt{R} in the package \texttt{hhh4underreporting}. The package can be installed from the repository \url{https://github.com/jbracher/hhh4underreporting} which also contains data and and code to replicate the presented application. Along with simulation codes, these are also available in the Supporting Information of this paper.

\end{document}